\documentstyle{article}

\title{{\bf Single-cluster algorithm for the site-bond-correlated Ising model}}

\author{P. R. A. Campos and R. N. Onody \\ \\
\small {\em Departamento de F\'{\i}sica e Inform\'{a}tica } \\
\small {\em Instituto de F\'{\i}sica de S\~{a}o Carlos} \\
\small {\em Universidade de S\~{a}o Paulo - Caixa Postal 369} \\
\small {\em 13560-970 - S\~{a}o Carlos, S\~{a}o Paulo, Brasil.}}
\date{}
\begin{document}
\maketitle
\normalsize
\baselineskip=24pt

\begin{abstract}

We extend the Wolff algorithm to include correlated spin interactions 
in magnetic systems. This algorithm is applied to study the
site-bond-correlated Ising model on a two dimensional square lattice.
We use a finite size scaling procedure to obtain the phase diagram in
the temperature-concentration ($T_{c},C$) space.
\\ \\ \\ \\ \\ 
\end{abstract}

\newpage

For randomly diluted magnetic systems the critical concentration where 
the magnetic order vanishes is of great interest. The dependence of the
critical temperature with the concentration is ruled by the topological 
properties of the lattice and by the symmetry of the interaction Hamiltonian.

In the simplest description of a dilute magnet, the occupied lattice sites
correspond to the magnetic atoms and the empty sites are associated with 
the presence of nonmagnetic atoms (impurities). Interactions are short ranged
usually of the exchange interaction type.
The strength of the exchange interaction is not affected by their neighbors.
At $T=0$, the cluster structure is well described by the ordinary site 
percolation model. 

However, it has been found experimentally that sometimes the {\em local
environment} may modify the exchange coupling constant between two atoms or
may even supress their magnetic manifestations. This means that interations
involving more than two atoms are present in the system. To take these
effects into account, two kinds of correlations were proposed: the long and
the short ranged correlation models. The bootstrap percolation model \cite
{cha} is a good example for the former and the site-bond-correlated model 
\cite{albi,albino} for the latter.
Long-ranged correlation may change the critical exponents or even the order
of the transition of its corresponding uncorrelated model. In contrast, the
short-ranged correlation seems to be unable to such drastic effects. Anyway,
for these systems, dilution as well as correlation play an important role.
 
An analysis of \hspace{0.1cm} $ ^{19}F $ \hspace{0.1cm} NMR linewidths in
the randomly diluted magnetic
system $ KNi_{x}Mg_{1-x}F_{3} $  and in the isostructural compound
$ KMn_{x}Mg_{1-x}F_{3} $ ( where $x$ is the concentration ) show remarkable 
differences in their properties \cite{albi,enge}. For instance, the 
concentration or percolation threshold where the magnetic order ceases are 
different. Further, the former displays an upward curvature in the 
temperature-concentration plane which is absent in the latter.
In order to explain these facts, Aguiar
{\em et al.} \cite{albi} proposed a dilution model where the exchange 
coupling constant of two ions $ Ni^{2+} $ depends directionally on the
magnetic attributes of theirs nearest neighbor atoms. That is the magnetic
spins are correlated. In a subsequent work \cite{albino}, a parameter 
$\alpha$ was introduced which is a measure of the correlation strength.
When $\alpha > 0$ or $\alpha < 0$ we have {\em ferro} or {\em
antiferromagnetic correlation}, respectively. Due to
the presence of both ferro and antiferromagnetic coupling constants, the
antiferromagnetic correlation brings competition to the system and a spin
glass behavior is expected.

The thermal properties of the site-bond-correlated Ising model 
(hereafter denoted SBC) was 
investigated using several approaches: mean field \cite{albino}, 
Honmura-Kaneyoshi effective-field \cite{silva}, renormalization group 
\cite{branco,branco2} and Monte Carlo renormalization group \cite{branco3}.
Any of these techniques lead to almost the same physical scenario with
the exception of the SBC defined on the Bethe lattice \cite{couti}.
Here there are some unexpected features which seem to be due to the
pathological geometry of the Bethe lattice.  
Also the SBC cluster characteristics has been studied and its connection
with the percolation problem. For the square lattice, it is now well
established that there are two kinds of percolations: the usual site
percolation with threshold $p_{c} \sim 0.592$ when $0 < \alpha \leq 1$ and 
a {\em correlated} percolation with threshold $p_{c} \sim 0.740$ at
$ \alpha=0$ \cite{bonfim}. There are now strong indications that both
percolations belong to the same universality class \cite{moura,campos}.

As far as we know, reference \cite{souza} is the only work which treats 
the thermal properties of the SBC model by using the Monte Carlo technique.
In that paper, the parameter $ \alpha $ is resctricted to be 
null. Everyone knows that is very difficult to simulate near
phase transitions due to the emergence of the critical slowing down
phenomenon. In the SBC this becomes even worst since dilution together
with correlation conspire to weaken the Metropolis technique.
Fortunately, for the so called {\em cluster algorithms} the dilution
turns these algorithms even more robust \cite{henne}. In the
Ising model (diluted or not) the Metropolis algorithm is a local 
Monte Carlo method where only one spin is flipped each time. For the 
{\em cluster algorithms} however, the entire cluster can be flipped. Indeed,
there are actually two kinds of cluster algorithms: the single-cluster
algorithm like the Wolff \cite{wolff} algorithm and those involving 
multiple clusters as the Swendsen-Wang \cite{swe} and the invaded cluster
\cite{mach} algorithms. In this paper we generalize the Wolff algorithm 
to include correlation. We use our results to construct the SBC phase
diagram for many concentrations and for the correlation $ \alpha $ in
the interval $[0,1]$.
    
In spite of the great interest received during the last decades by the 
critical slowing down phenomenon, it is still a hard question to
find under what conditions it can be reduced. Two independent strand of
thought had contribute to enlighten the problem. The first was the search
for the so called "physical cluster" of the Ising model. The idea is to 
find a percolation problem in absolute harmony with the thermal problem:
the temperature where the percolation threshold occurs ($T_{p}$) must
coincide with the critical temperature ($T_{c}$) and also their
corresponding critical exponents need to be the same. The "geometrical
cluster", i. e., those clusters formed by grouping all nearest neighbors
up (or down) spins were discarded since in 3-d they neither have
$T_{p}=T_{c}$ nor are their exponents the same. The first explicit
construction of "physical clusters" was proposed by Coniglio and Klein
\cite{coni}. Parallel to this research line, Fortuin and Kasteleyn
\cite{fort} introduced the random cluster model. They proved that the
susceptibility of the Ising model is equal to the mean cluster size of
the random cluster model. Gathering these ideas, Swendsen and Wang 
\cite{swe} developed
a new algorithm for the Potts model. In this algorithm, clusters of
spins in the same states are grown and flipped. Thus, a number of spins
are update in a single move and the correlation time is reduced.

In 1989, a {\em single-cluster} Monte Carlo algorithm was introduced by
Wolff \cite{wolff} for the {\it O(N)} spin models as a variation on the 
Swendsen-Wang scheme for the Ising model and has proved to be even more
effective. In the Wolff algorithm only one cluster is formed and flipped
with probability 1, whereas in the Swendsen-Wang dynamics all percolation
clusters are formed and flipped with probability 1/2. According to Tamayo
et al. \cite{tama} the main reason for a better performance of the Wolff 
algorithm is that the mean size of the clusters flipped is significantly
larger than in Swendsen-Wang case.

In the SBC model the presence of nonmagnetic impurities in the
neighborhood of a given pair of nearest-neighbor magnetic atoms can modify
the strength of the exchange interaction between the two atoms. Moreover,
in the limit of strong correlation $\alpha$, the correlation can even
suppress the relevant exchange interaction.

The model Hamiltonian is the following

\begin{equation}
H=-\sum_{i,\delta }J_{i,i+\delta }(\sigma _{i}\sigma _{i+\delta }-1)
\end{equation}
where $\sigma _{i}=\pm 1$ and $\delta $ denotes an elementary lattice
vector. The exchange interaction $J_{i,i+\delta }$ is given by

\begin{equation}
J_{i,i+\delta }=J\varepsilon _{i}\varepsilon _{i+\delta }\left[ \left(
1-\alpha \right) \varepsilon _{i-\delta }\varepsilon _{i+2\delta }+\alpha
\right] ,
\end{equation}
where $J>0$. The random variables $ \varepsilon _{i}$ can take values one
with probability $C$ and zero with probability $1-C$, where $C$ is the
concentration of magnetic atoms. The parameter $\alpha $ correlates the
interaction between sites $i$ and $i+\delta $ with the magnetic occupancy
of the sites $i-\delta $ and $i+2\delta$. The uncorrelated dilute Ising model
is re-obtained in the limit $\alpha =1.$ For $0<\alpha <1$, the bond between
$i$ and $i+\delta $ is only weakened by the absence of a magnetic atom at
$i-\delta $ or $i+2\delta .\,$ The limit $\alpha =0$ corresponds to the
maximum correlation, i. e., two magnetic first neighbor sites are
connected by an active bond only if their nearest-neighbor sites along
the line joining them are also present. 

In the Wolff algorithm for the pure Ising model, first a site is randomly
chosen in the lattice. A nearest neighbour site will be added to the cluster
with an activation probability $p=1-e^{-2K}$ (where $K = \frac{J}{K_{B}T}$ )
if it is in the same spin state. This procedure is repeated until no more
sites can be incorporated to the cluster. The whole cluster is then flipped.
Following Fortuin-Kasteleyn \cite{fort}, we derive the bond activation
probability: 

\vspace{0.6cm}

$\left\{ 
\begin{tabular}{l}

$p_{i.i+\delta }=0 $ \ \ \ \ \ \ \ \ \ \ \ \ \ \ \ \ \ if \ $\sigma _{i} \neq 
\sigma _{i+\delta}$ \\
$p_{i.i+\delta }=1-e^{-2K} $ \ \ \ \ \ \ \ if \ $\sigma _{i}=\sigma _{i+\delta
}$ \ and \ $\varepsilon _{i-\delta }\varepsilon _{i+2\delta }=1$ \\ 
$p_{i.i+\delta }=1-e^{-2\alpha K} $ \ \ \ \ \ if \ $\sigma _{i}=\sigma
_{i+\delta }$ \ and \ $\varepsilon _{i-\delta }\varepsilon _{i+2\delta }=0$%

\end{tabular}
\right. $ 

\vspace{0.6cm}

\noindent between sites $i$ and $i+\delta$.
The definition of these activation probabilities guarantees correctly the
applicability of the cluster dynamics and constitutes the Fortuin-Kasteleyn
mapping for the SBC model.

We simulate the SBC model for various values of correlation $\left( \alpha 
\right) $ and concentration $\left( C\right)$ on a square lattice of size
$ L=50,100,150$ and $300$. A random uniformly distributed magnetic sites
configuration (or simply {\em sample}) is generated with an occupation
probability $ C $. Over this quenched (geometric) configuration, an
{\em initial} spin configuration is chosen with half of the spins up.
One Wolff's cluster is then constructed using the activation probabilities
just described. The entire cluster is then flipped and the new magnetization
is determined. We call the sequence: cluster construction + flipping +
magnetization measure {\em one} iteration or Monte Carlo step. Our first
1000 iterations were discarded waiting the system to achieve the thermal
equilibrium. The remaining iterations were used to measure the mean
magnetization and its fluctuation - the magnetic susceptibility. On these
quantities, another average over the samples was necessary in order to
anneal the geometric influence (always present in diluted systems).
Finally, we scan the coupling $K$ to find $K_{max}$ where the susceptibility
is maximum. In Table 1 we show the number of iterations and realizations
necessaries to reduce the error of $K_{max}$ to $\sim 0.01$. Of course, they
depend on the values of $\alpha$ and $C$.

\begin{center}
Table 1 to be inserted
\end{center}

The determination of $K_{max}$ was done for fixed values of $\alpha$, $C$
and $L$ and then extrapolated to the thermodynamic limit $L \rightarrow 
\infty$ through the BST algorithm \cite{bst}. The BST is a useful algorithm
to extrapolate physical quantities that converge obeying a power law $%
F(L)=F(L=\infty )+AL^{-\Theta }$. It allows a reliable determination of
critical parameters in the limit $L \rightarrow \infty $ and its versatility
becomes more pronounced if there are only very short sequences available.

\begin{center}
Figure 1 to be inserted
\end{center}

The phase diagram is shown in Fig. 1. The initial slope $(1/T_{c})(dT_{c}/dC)$ at
the pure $C=1$ case increases with the degree of correlation, which is 
supported by both experimental \cite{albi} and theoretical \cite{albino,silva,
souza} data. At $T=0$, we obtain two distinct percolation threshold: one
for $\alpha = 0$ and another for $ \alpha \neq 0$. It can be seen in the
figure that the curves for intermediate correlation $ 0< \alpha < 1$ have 
an upward curvature ( not present in extreme the cases $\alpha =0$ and
$\alpha =1$) in agreement with the experimental results for the 
$ KNi_{x}Mg_{1-x}F_{3} $ compound \cite{albi}. In summary, we have 
generalized the Wolff's algorithm for systems where both dilution and correlation
are present. This allowed us to get the phase diagram of the site-bond-correlated
Ising model on a two dimensional square lattice. The natural question on how
the new algorithm changes the dynamical critical exponent and the autocorrelation
time is now under investigation.

We acknowledge Conselho Nacional de Desenvolvimento Cient\'{\i}fico e Tecnol\'ogico
(CNPq) and Funda\c c\~ao de Amparo a Pesquisa do Estado de S\~ao Paulo (FAPESP) for 
financial support.

\newpage

\begin{center}
{\bf Table Caption}
\end{center}

\vspace{1.5cm}

{\bf Table 1} - Range of the number of samples $N_{s}$ and number of iterations
$N_{i}$ used in this work. The numbers $N_{s}$ and $N_{i}$ augment for 
smalls $\alpha$ (strong correlation) and $C$ (strong dilution).

\newpage

\begin{center}
{\bf Figure Caption}
\end{center}

\vspace{1.5cm}

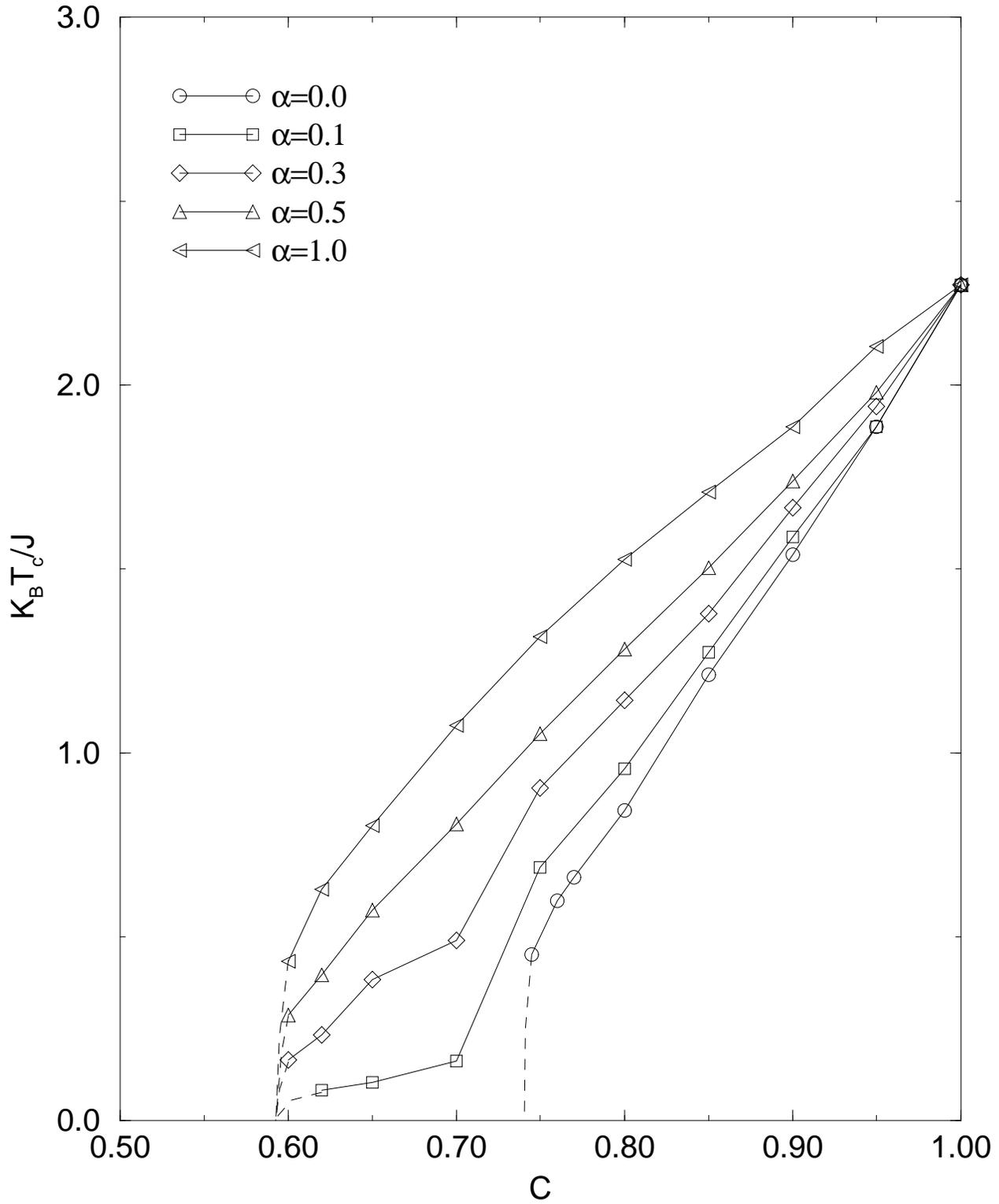
{\bf Fig. 1} - The SBC Ising model phase diagram for many values of $\alpha$.
The percolations threshold are $0.740$ and $0.592$ for $\alpha=0$ and 
$ 0< \alpha \leq 1 $, respectively. The dashed lines are only a guide to
the eye.

\newpage

\begin{table}
\begin{center}
\begin{tabular}{||c|c|c||}
\hline\hline
\multicolumn{1}{||c|}{\bf $L$}&
\multicolumn{1}{c|}{\bf $N_{s}$}&
\multicolumn{1}{c||}{\bf $N_{i}$} \\
\hline
$50$ & $20-150$ & $3000-5000$ \\ 
\hline
$100$ & $05-50$ & $3000-5000$ \\ 
\hline
$150$ & $05-50$ & $3000-5000$ \\ 
\hline
$300$ & $05-50$ & $3000-10000$ \\ 
\hline\hline
\end{tabular}
\vspace{1.5cm}
\end{center}
\end{table}
\begin{center}
Table 1
\end{center}


\newpage

\begin{thebibliography}{99}

\bibitem{cha} J. Chalupa, P. L. Leath and G. R. Reich, J. Phys. C {\bf 12},
L31 (1979).

\bibitem{albi} J. Albino O. de Aguiar, M. Engelsberg and H. J. Guggenheim,
J. Magn. Magn. Mater.{\bf 54-57},107 (1986).

\bibitem{albino} J. Albino O. de Aguiar, F. G. Brady Moreira and M. 
Engelsberg, Phys. Rev. B {\bf 33}, 652 (1986).

\bibitem{enge} M. Engelsberg, J. Albino O. de Aguiar, O. F. de Alcantara
Bonfim and A. Franco Jr., Phys. Rev. B {\bf 32}, 7143 (1985).
  
\bibitem{silva} A. A. P. da Silva and F. G. B. Moreira, Phys. Rev. B
{\bf 40}, 10986 (1989).

\bibitem{branco} N. S. Branco, S. L. A. de Queiroz and R. R. dos Santos,
Phys. Rev. B {\bf 38}, 946 (1988).

\bibitem{branco2} N. S. Branco, S. L. A. de Queiroz and R. R. dos Santos,
Phys. Rev. B {\bf 42}, 458 (1990).

\bibitem{branco3} N. S. Branco and K. D. Machado, Phys. Rev. B {\bf 47},
493 (1993).

\bibitem{couti} S. Coutinho, J. A. O. de Aguiar, F. G. B. Moreira and 
J. R. L. de Almeida, Phys. Rev. B {\bf 36}, 8478 (1987).

\bibitem{bonfim} O. F. A. Bonfim and M. Engelsberg, Phys. Rev. B {\bf 34},
1977 (1986).

\bibitem{moura} L. M. de Moura and R. R. dos Santos, Phys. Rev. B
{\bf 45}, 1023 (1992).

\bibitem{campos} P. R. A. Campos, L. F. C. Pessoa and F. G. B. Moreira,
to be published in Phys. Rev. B.
 
\bibitem{souza} A. J. F. de Souza and F. G. B. Moreira, J. Phys. (Paris)
{\bf 49}, C8 (1988).

\bibitem{henne} M. Hennecke and U. Heyken, J. Stat. Phys. {\bf 72},
829 (1993).

\bibitem{wolff} U. Wolff, Phys. Rev. Lett. {\bf 62}, 361 (1989).

\bibitem{swe} R. H. Swendsen and J. S. Wang, Phys. Rev. Lett. {\bf 58},
86 (1987).

\bibitem{mach} J. Machta, Y. S. Choi, A. Lucke and T. Schweizer, Phys.
Rev. Lett. {\bf 75}, 2792 (1995).

\bibitem{coni} A. Coniglio and W. Klein, J. Phys. A {\bf 13}, 2775
(1980).

\bibitem{fort} P. W. Kasteleyn and C. M. Fortuin, J. Phys. Soc. Japan
{\bf 26} (Suppl.), 11 (1969); C. M. Fortuin and P. W. Kasteleyn,
Physica {\bf 57}, 536 (1972). 

\bibitem{tama} P. Tamayo, R. C. Brower and W. Klein, J. Stat. Phys.
{\bf 58}, 1083 (1990).

\bibitem{bst} R. Bulirsch and J. Stoer, Numer. Math. {\bf 6}, 413 (1964).

\end{thebibliography}
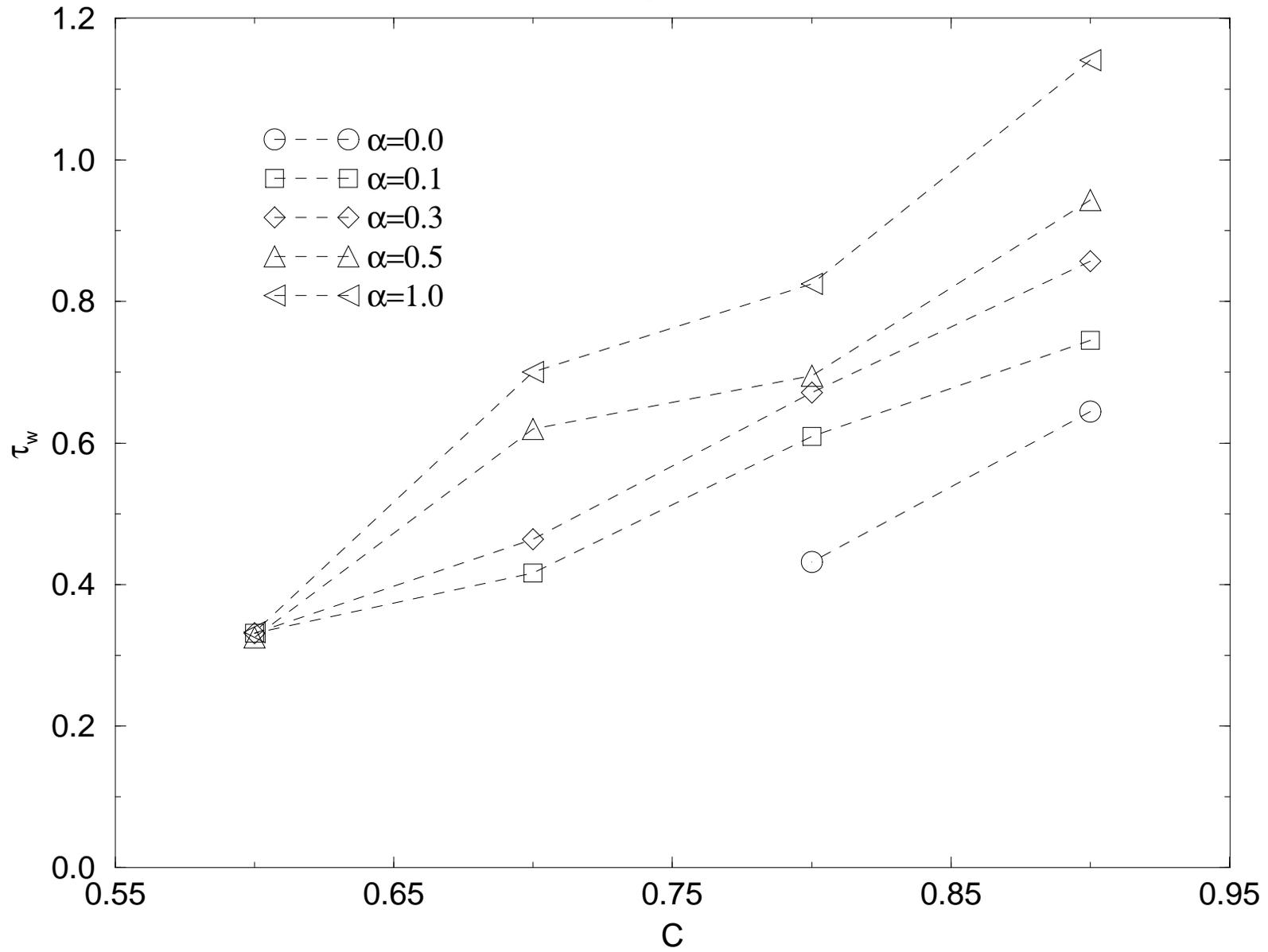
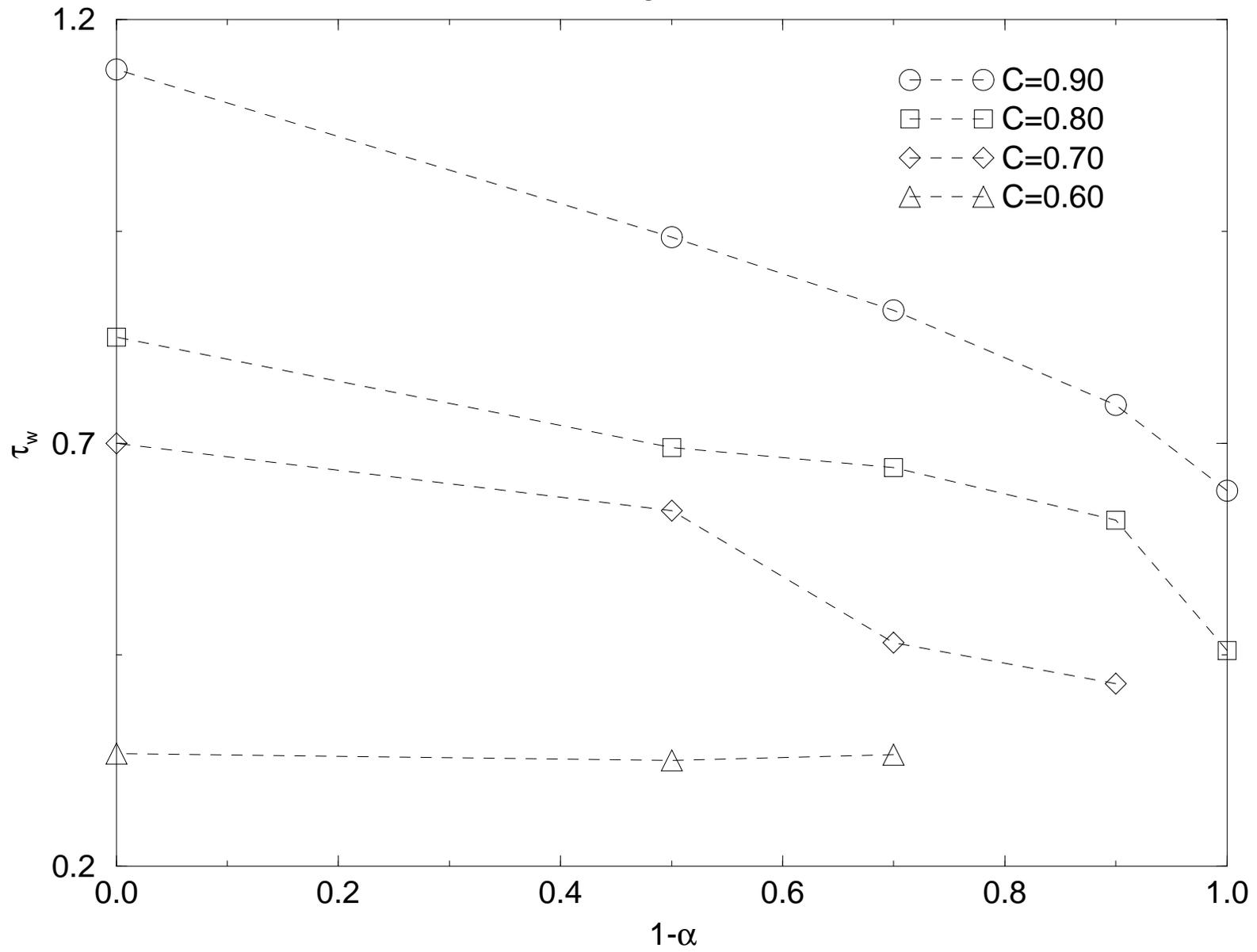
\end{document}